\documentclass[doublecol]{epl2} % for 2 columns style with line numbers
\usepackage{amssymb}

\newcommand{\be}{\begin{equation}}
\newcommand{\ee}{\end{equation}}
\newcommand{\bd}{\begin{displaymath}}
\newcommand{\ed}{\end{displaymath}}
\newcommand{\BE}{\begin{eqnarray}}
\newcommand{\EE}{\end{eqnarray}}

\newcommand{\erf}{{\rm erf}}

\newcommand{\bx}{\ensuremath{\mathbf{x}}}

\newcommand{\boldpsi}{{\mbox{\boldmath $\psi$}}}

\newcommand{\avg}[1]{\left\langle{#1}\right\rangle}

\usepackage{color}

\title{Dynamically evolved community size and stability of random Lotka-Volterra ecosystems}
\shorttitle{Evolved community size and stability of random Lotka-Volterra systems} %Insert here a short version of the title if it exceeds 70 characters

\author{Tobias Galla}

\institute{                    
  Theoretical Physics, School of Physics and Astronomy, The University of Manchester, Manchester M13 9PL, United Kingdom

}

\abstract{We use dynamical generating functionals to study the stability and size of communities evolving in Lotka-Volterra systems with random interaction coefficients. The size of the eco-system is not set from the beginning. Instead, we start from a set of possible species, which may undergo extinction. How many species survive depends on the properties of the interaction matrix; the size of the resulting food web at stationarity is a property of the system itself in our model, and not a control parameter as in most studies based on random matrix theory. We find that prey-predator relations enhance stability, and that variability of species interactions promotes instability. Complexity of inter-species couplings leads to reduced sizes of ecological communities. Dynamically evolved community size and stability are hence positively correlated. }

\begin{document}

\maketitle

\section{Introduction}

One of the most controversial debates in ecology concerns the question
whether complexity of species communities begets stability. Early
studies suggested
that densely connected food webs can cope better with the loss of a
single link or an external perturbation than poorly interwoven
networks with only a small number of links or energy flow
pathways  \cite{macarthur, elton}. Theoretical analyzes by Gardner and Ashby \cite{gardner}
 and May \cite{may1,may2} in the 1970s however suggested that
 complex community models may not always be more stable than less diverse ones. But `if
increased diversity does not necessarily result in greater stability',
as Rooney et al put it \cite{rooney}, then
`why do diverse food webs seem to be more stable than depauperate
ones' in ecological field studies? Advances in this so-called
diversity-stability debate are often, as in Ashby's and Gardner's and
in May's work, based on random community models, in which the
coefficients describing interactions between species are drawn at
random \cite{mccann}. Species are then often assumed to follow a community dynamics
described e.g. by Lotka-Volterra or replicator equations \cite{hofbauer, nowak}. The issue of stability versus complexity is then
addressed by studying the properties of fixed-points of these dynamics
in dependence on model parameters such as the mean interaction
strength, their variance, the mean connectance and the size of the
community under consideration.

In the present work we use methods from statistical mechanics and
the theoretical physics of disordered systems \cite{mpv, dominicis, msr, coolendyn,coolenbook1,coolenbook}
combined with concepts of non-linear dynamics and the theory of
stochastic processes to study random community models
and to make mathematically exact predictions regarding the stability
or otherwise of their dynamics. Conceptually this is similar to May's
approach \cite{may1,may2} in that we address models with random
interaction matrices. Recent work on random communities includes \cite{allesina2012, tang, allesina2015, grilli, gibbs}.

 The difference of our approach compared to this existing work is as follows. Much work on random community Lotka-Volterra models in the ecological
literature is concerned with eco-systems of a given pre-arranged fixed
size. It then assumes a random Jacobian matrix of that size, and uses random matrix theory to study the eigenvalues of these matrices \cite{gardner, may1, may2, rozdilsky, jansen,allesina2012, tang, allesina2015,grilli}. In many cases no actual dynamics are specified -- the starting point is the Jacobian of the surviving species. In our approach, the size of the resulting eco-system is not set from the beginning. Instead, we start from a set of possible species, specify its dynamics (Lotka-Volterra) and assume random interaction coefficients. The species in our model may undergo
extinction and hence some species will not survive in the long-term
limit. How many species go extinct or survive depends on the properties of the interaction matrix. Crucially, the size of the resulting food web at stationarity is  a property of the system itself in our model, and not a
control parameter as in most studies based on random matrix theory. 

Furthermore, the bulk of the existing
literature on random community models in theoretical ecology is
restricted to stability analyses. Stable and unstable regimes are identified from the application of random matrix theory to presumed random Jacobians, but only few statements are made about the properties of stable fixed points. As part of our study, we also pursue a linear
stability analysis and investigate in detail how e.g. the presence of
predator-prey pairs in the community affect the stability of the
underlying dynamics. But the techniques we use allow us also to calculate fixed-point properties and the statistics of the ecological community at stationarity. In particular we obtain results for species abundance
and rank distributions, the fraction of surviving species and the
total biomass contained in the system. No approximations need to be
made (except for assuming the community under consideration to be
large). Our theoretical predictions are confirmed convincingly in
numerical simulations.

Our work build on a a number of existing studies. In the statistical physics community replicator models with random
couplings have first been proposed by Opper and Diederich \cite{diederich, opper}, and stable and unstable
regimes of such model systems have been identified and characterised
analytically within the theory of phase transitions of statistical
mechanics, see also \cite{biscari, fontanari, tokita1, tokita2, gallaasym,gallahebb, yoshino1, yoshino2}.  Similar tools can also be used to study game learning \cite{gallafarmer} and the distribution of Nash equilibria in games \cite{berg1,berg2, gallaepl}. 

These existing non-equilibrium statistical physics studies of random community models are restricted to replicator models in which the total concentration of species is conserved. Furthermore, results have been expressed mostly in dependence on a so-called co-operation
pressure; an intra-species interaction term suppressing the growth of
individual species, and driving the system to a state of diversity. In the present paper we
address Lotka-Volterra systems and focus the effects of complexity and
variability on the level of inter-species interactions and address
questions of feasibilty as well. While a formal mathematical
equivalence between replicator systems and Lotka-Volterra systems (of
a different dimensionality) can be established (see e.g. \cite{hofbauer}) replicator systems are inherently bounded by definition,
and do not allow for runaway solutions. In
Lotka-Volterra systems, on the contrary, the total biomass is a
dynamical quantity, and can be computed analytically from the
statistical physics theory. Furthermore, as we will see below,
Lotka-Volterra systems show an instability, distinctly different from
that of replicator systems, separating stable fixed-point regimes
from phases in which characteristic quantities such as the biomass and
individual species concentrations diverge in time. No such regime is
found in random replicator systems, where instead bounded and
potentially chaotic trajectories are observed in the unstable regime
\cite{opper,gallaasym, gallahebb}.

\section{Lotka-Volterra random community model}
We consider a generalized Lotka-Volterra model describing the dynamics of
an interacting community of $N$ species, labeled by $i=1,\dots,N$. The time-dependent number density of individuals of species $i$ is denoted by $x_i(t)$, and evolves in
time according to
\be\label{eq:lv}
\frac{dx_i(t)}{dt}=r_ix_i(t)\left(K_i+\sum_{j=1}^N \alpha_{ij} x_j(t)\right).
\ee
The intra-specific
interaction coefficients $\alpha_{ii}$ will be set to $\alpha_{ii}=-1$, following for example \cite{may1, jansen}. For simplicity, we set the basic growth rates $r_i$ to unity. The quantities $K_i$ denote carrying capacities; if there are no interactions between species ($\alpha_{ij}=0$ for $i\neq j$) then $\dot x_i=x_i (K_i-x_i)$. We focus on the case $K_i=1$ for all $i$. The interaction coefficients $\alpha_{ij}$ ($i\neq j$) finally represent the (per capita) effect of species on one another. A negative
coefficient $\alpha_{ij}$  indicates a competitive effect of species $j$ on species $i$.  

In our setup the couplings $\alpha_{ij}$ ($i\neq j$) are drawn from a Gaussian random distribution \cite{may1,may2, diederich, opper, gallaasym}  characterized by its mean and covariance matrix. We introduce a model parameter controlling the correlation between the interaction coefficients $\alpha_{ij}$ and $\alpha_{ji}$, and hence the fraction of prey-predator pairs in the artificial ecological community. A prey-predator pair consists of two species $i$ and $j$ for which $\alpha_{ij}$ and $\alpha_{ji}$ have opposite signs, i.e. a pair in which the presence of say species $i$ has a detrimental effect on species $j$, whereas the presence of species $j$ is beneficial for individuals of species $i$, see also \cite{allesina2012}.

Specifically for any pair $i<j$ of species we set
\be
\alpha_{ij}=\frac{\mu}{N}+\frac{\sigma}{\sqrt{N}}z_{ij}, ~~~ \alpha_{ji}=\frac{\mu}{N}+\frac{\sigma}{\sqrt{N}}z_{ji}, 
\ee
where $z_{ij}$ and $z_{ji}$ are drawn from a Gaussian distribution with $\overline{z_{ij}}=0$, $\overline{z_{ij}^2}=1$, and $\overline{z_{ij}z_{ji}}=\gamma$. The overbar describes averages over the Gaussian ensemble. The scaling of the moments of the $\alpha_{ij}$ with $N$ is necessary to produce a well defined limit $N\to\infty$ in which the statistical mechanics theory applies. The parameter $-1\leq\gamma\leq 1$ characterizes the correlations between $z_{ij}$ and $z_{ji}$. For $\gamma=1$ one has $\alpha_{ij}=\alpha_{ji}$ with probability one. For $\gamma=0$, $z_{ij}$ and $z_{ji}$ are uncorrelated, and for $\gamma=-1$ one has  $z_{ij}=-z_{ji}$ with probability one. In the limit of large system size, $N\to\infty$, a given pair of species $i\neq j$ form a predator-prey pair ($\alpha_{ij}\alpha_{ji}<0$) if and only if $z_{ij}$ and $z_{ji}$ are of opposite sign. The percentage $p$ of predator-prey interactions can hence be worked put by performing a suitable Gaussian integral over the joint distribution of $z_{ij}$ and $z_{ji}$. This leads to an explicit, non-linear and decreasing dependence of $p$ on $\gamma$. In particular one has $p=1$ for $\gamma=-1$ (for $\gamma=-1$ the system consists fully of predator-prey interactions and in the limit of large $N$); one has $p=1/2$ for $\gamma=0$ ($50\%$ predator-prey pairs), and $p=0$ for $\gamma=1$ (i.e. no prey-predator pairs are present for $\gamma=1$). In all cases, the remaining fraction of $1-p$ interaction pairs are non-predator-prey. In the limit $N\to\infty$, half of these will be of a mutualistic interaction type ($\alpha_{ij}$ and $\alpha_{ji}$ both positive), and the other half of a strictly competitive type ($\alpha_{ij}$ and $\alpha_{ji}$ both negative).\\

%Similar systems have been analyzed in Jansen \& Kokkoris 2003, Rozdilsky \& Stone 2001 and by Wilson et al. (Wilson, Lundberg et al 2003). The work by Jansen et. al. and by Rozdilsky and Stone is mostly based on computer simulations, and the analytical results obtained by Wilson et al. are mostly of an approximate nature (albeit with good agreement with measurements in computer simulations).  

\section{Path-integral analysis}

We study the random community Lotka-Volterra model,
Eq. (\ref{eq:lv}), in the limit of a large number of interacting
species ($N\to\infty$) using dynamical methods from spin-glass physics \cite{dominicis, msr, mpv, coolendyn,coolenbook}.

The starting point of the path-integral analysis are the $N$-species Lotka-Volterra equations
\be\label{eq:lv2}
\frac{dx_i(t)}{dt}=r_ix_i(t)\left(K_i+\sum_{j=1}^N \alpha_{ij} x_j+h(t)\right),
\ee
where we have added a perturbation field $h(t)$, which will be used to
generate dynamical response functions and susceptibilities. This field
is a theoretical device and is set to zero at the end of
the calculation. The dynamical moment generating functional is given by 
\BE
Z[\boldpsi]&=&\int [D\mathbf{x}] \prod_{it} \exp\left(i\sum_i\int dt \,\psi_i(t)x_i(t)\right) \nonumber \\
&&\times\delta\bigg(\dot x_i-x_i(t)[1+\sum_{j=1}^N \alpha_{ij} x_j+h(t)]\bigg),
\EE
where $\delta(\cdots)$ denotes the (functional) Dirac delta-distribution, and restricts the integral to all paths allowed by the Lotka-Volterra dynamics. The notation $[D\mathbf{x}]$ indicates a functional integral over trajectories of the system. The variables $\psi_i(t)$ represent external source fields; $Z[\boldpsi]$ hence describes the (functional) Fourier transform of the measure generated by the Lotka-Volterra dynamics in the space of possible trajectories. Performing the average over all possible realizations of interaction matrix entries $\{\alpha_{ij}\}$ along the lines of \cite{diederich, opper, gallaasym} leads, in the limit $N\to\infty$, to the following stochastic process for the concentration $x(t)$ of a representative species
\BE
\frac{dx(t)}{dt}&=&x(t)\bigg[1-x(t)+\mu M(t)\nonumber \\
&&+\gamma\sigma^2\int_{0}^t G(t,t')x(t')dt'+\eta(t)+h(t)\bigg], \label{eq:effective}
\EE
see also the Supplementary Material. We now describe the different ingredients of this process. We have coloured Gaussian noise $\eta(t)$, with temporal correlations given self-consistently by $\avg{\eta(t)\eta(t')}=\sigma^2\avg{x(t)x(t')}$, where $\avg{\dots}$ denotes an average over the process in Eq. (\ref{eq:effective}).  The effective-species concentration $x(t)$ thus is a random process itself. A further component of the effective dynamics is the non-Markovian term coupling back in time through the integral over $t'$. This term and the coloured noise $\eta$ are remnants of the initial randomness of the species interactions $\{\alpha_{ij}\}$. The key quantities describing the dynamics of the model are the correlation function $C(t,t')$, the response function $G(t,t')$ and the average species concentration $M(t)$, or equivalently the total biomass in the system at time $t$. These order parameters are to be obtained self-consistently as averages over realizations of the effective-species process as
\BE
C(t,t')&=&\avg{x(t)x(t')},\nonumber \\
G(t,t')&=&\avg{\frac{\delta x(t)}{\delta h(t')}}, \nonumber \\
M(t)&=&\avg{x(t)}.
\EE
A fixed-point ansatz $\lim_{t\to\infty}x(t)= x^*$, $\lim_{t\to\infty}\eta(t)=\eta^*$, with both $x^*$ and $\eta^*$ static random variables then leads to $M(t)\equiv M$, and $C(t,t')\equiv \sigma^2 q$, where $q=\avg{(x^*)^2}$. This is similar to the procedure in \cite{opper, gallaasym}. Within this fixed-point ansatz $G(t,t')$ becomes time-translation invariant, i.e., $G(t,t')$ is a function only of $\tau=t-t'$. Causality dictates $G(\tau)=0$ for $\tau<0$. We write $\chi=\int_0^\infty d\tau~G(\tau)$. This ansatz leads to
\be
x^*\left[1-x^*+\mu M+\gamma\sigma^2\chi x^*+\eta^*\right]=0,
\ee
so that fixed points can take values $x^*=0$ and $x^*=(1+\mu M+\eta^*)/(1-\gamma\sigma^2\chi)$. The latter solution is only physical if it is non-negative, so that we have

\be
x^*(\eta^*)= \frac{1+\mu M+\eta^*}{1-\gamma\sigma^2\chi}H\left(\frac{1+\mu M+\eta^*}{1-\gamma\sigma^2\chi}\right),
\ee
where $H(x)$ is the Heaviside function, $H(x)=1$ for $x>0$, and $H(x)=0$ else. Note that $\eta^*$ is a Gaussian random variable, as indicated above, so $x^*$ is a random quantity as well. These results re-iterate that a fraction of the $N$ initial species  dies out during the transients of the Lotka-Volterra dynamics, and are no longer present at the fixed points.

Following the lines of \cite{opper, gallaasym} to perform the average over the ensemble of fixed points one finds closed non-linear integral equations (see Supplementary Material),
\BE
\chi&=&\frac{1}{1-\gamma\sigma^2\chi}\int_{-\infty}^\Delta \frac{dz}{\sqrt{2\pi}}e^{-z^2/2},\label{eq:fp1}\\
M&=&\frac{\sqrt{q}\sigma}{1-\gamma\sigma^2\chi}\int_{-\infty}^\Delta\frac{dz}{\sqrt{2\pi}}e^{-z^2/2}(\Delta-z),\label{eq:fp2}\\
1&=&\frac{\sigma^2}{(1-\gamma\sigma^2\chi)^2}\int_{-\infty}^\Delta\frac{dz}{\sqrt{2\pi}}e^{-z^2/2}(\Delta-z)^2,\label{eq:fp3}
\EE
for $q, M$ and the dynamic susceptibility $\chi$. We have used the abbreviation
$\Delta=(1+\mu M)/(\sqrt{q}\sigma)$. The nature of the fixed point ansatz is such that it disregards any dependence on initial conditions. Similar to \cite{opper, gallaasym} it applies under the assumption that the initial Lotka-Volterra equations (\ref{eq:lv}) have a stable fixed point, and that his fixed point is unique for any given realisation of the $\{\alpha_{ij}\}$, see the Supplement for further details.

Equations (\ref{eq:fp1},\ref{eq:fp2},\ref{eq:fp3}) can be
solved by Newton-Raphson methods, and deliver $q,\chi$ and $M$ as a
function of the model parameters $\sigma,\mu$ and $\gamma$. The
fraction of surviving species is obtained from these solutions as
$\phi=\frac{1}{2}\left(1+\erf(\Delta/\sqrt{2})\right)$.

\section{Stability and phase diagram}

A linear stability analysis of the fixed point solution can be performed
along the lines of \cite{opper}. We here only summarise this briefly, and relegate details to the Supplement. One starts from
\BE
\frac{dy(t)}{dt}&=&x^*\left[-y(t)+\gamma\sigma^2\int_{0}^t G(t,t')y(t')dt'\right.\nonumber \\
&&\left.+v(t)+\xi(t)\right], \label{eq:pert}
\EE
where $y(t)$ denotes fluctuations about a (non-zero) fixed point $x^*$, and where $v(t)$ is the corresponding deviation in the noise in the effective process. The quantity $\xi(t)$ is Gaussian white noise of unit amplitude, generating the fluctuations about the fixed point. Self-consistently one has $\avg{v(t)v(t')}=\sigma^2\avg{y(t)y(t')}$. One converts into Fourier space and obtains
\BE
\frac{i\omega\widetilde y(\omega)}{x^*}&=&\left(\gamma\sigma^2 \widetilde G(\omega)-1\right) \widetilde y(\omega)+\widetilde v(\omega)+\widetilde \xi(\omega). \label{eq:pert2}
\EE
Focusing on long-time behaviour (i.e., $\omega=0$) leads to
\BE
\avg{|\widetilde y(0)|^2}&=& \phi \left[\gamma\sigma^2 \chi-1\right]^{-2}\left[\sigma^2\avg{|\widetilde y(0)|^2} +1\right]. \label{eq:pert3}
\EE
From this one finds that $\avg{|\widetilde y(0)|^2}$ diverges when $\phi\sigma^2=(1-\gamma\sigma^2\chi)^2$. This then leads to $\Delta=0$ in Eqs. (\ref{eq:fp1}, \ref{eq:fp2}, \ref{eq:fp3}), i.e. in particular $\phi=1/2$. From this we find that the stable fixed point ansatz is valid (in the sense that perturbations about it do not diverge) for $\sigma<\sigma_c$, where $\sigma_c$ depends on $\gamma$ via
\be\label{eq:stab}
\sigma_c^2(\gamma)=\frac{2}{(1+\gamma)^2}.
\ee
The theory, based on a stability assumption, hence self-consistently predicts its own breakdown, and the onset of instability. In the regime $\sigma<\sigma_c$ we therefore expect the original Lotka-Volterra dynamics to have stable fixed points in the limit of large $N$. One obtains $\sigma_c^2=0.5$ for $\gamma=1$ (no predator-prey pairs), $\sigma_c^2=2$ for $\gamma=0$ ($50\%$ predator-prey pairs), and $\sigma_c=\infty$ for $\gamma=-1$. In this latter case, in which all species pairs are of the predator-prey type, the system is thus predicted to be stable at any finite variance $\sigma^2$ of interaction strengths.  \\
\begin{figure}[t]
\begin{center}
\includegraphics[width=0.25\textwidth]{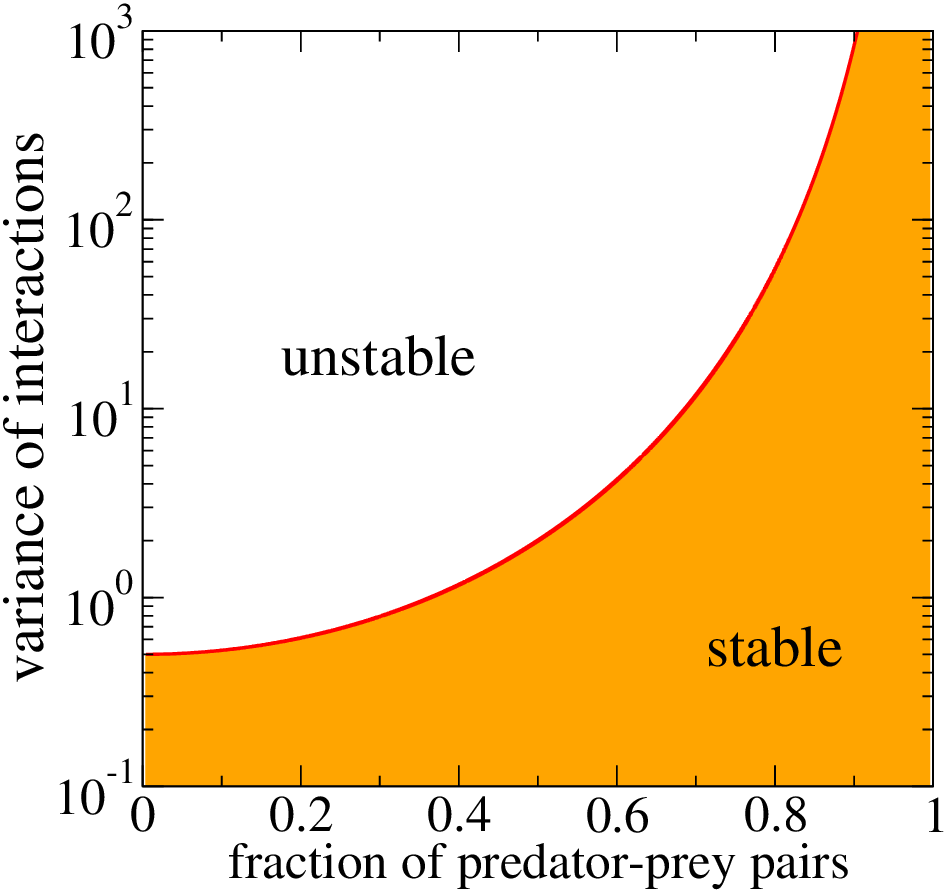}
\end{center}
 
\caption{\label{fig:pg} (Color on-line) 
Phase diagram. The critical standard deviation $\sigma_c^2$ of
interaction strengths obtained from the statistical physics
analysis as a function of relative frequency of predator-prey pairs at
the start of the dynamics.}
\end{figure}

The resulting phase diagram is shown in Fig. \ref{fig:pg}. We identify two different regimes of the Lotka-Volterra system: one
stable phase with a unique fixed point of the dynamics for variances
of the couplings strengths smaller than a threshold value
$\sigma_c^2$, and an unstable phase in which the total biomass
produced by the dynamics tends to infinity in the long run for large
variability in the interaction matrix ($\sigma^2>\sigma_c^2$). Our
analysis hence up to this point confirms the findings of May \cite{may1}, but allows for further analysis of the effects of the
interaction matrix on the stability properties of the ecological
network. As seen in Fig. \ref{fig:pg}, the threshold value
$\sigma_c^2$ depends on the correlation structure of the interaction
matrix, in particular an increased fraction of predator-prey pairs
leads to an increase in $\sigma_c^2$, i.e. the presence of
predator-prey pairs promote stability, in line with finding from random matrix theory \cite{allesina2012}. Indeed we find that the
threshold value $\sigma_c^2$ tends to infinity if all 
interaction pairs in the system are of the predator-prey type, and
that the eco-system is stable irrespective of the variance of
interactions in this case. Our theory thus supports e.g. the findings
by Bascompte et al. \cite{bascompte} and suggests
that predator-prey pairs and asymmetric interaction may be crucial for
the stability and maintenance of ecological communities. Our results
are also in-line with \cite{allesina} 
who studied random community models in which the interaction matrices
contain only predator-prey pairs. Their
computer simulations show that `when the interaction between species
is constrained to consumer-resource relationships, large and very
interconnected communities exhibit a high probability of stability
compared to the random case' \cite{allesina}, and that the region in parameter space
in which stability is likely `grows dramatically' when the relation
between species is constrained to be predator-prey.  

\section{Community properties}
\begin{figure}[t]
\begin{center}
\includegraphics[width=0.22\textwidth]{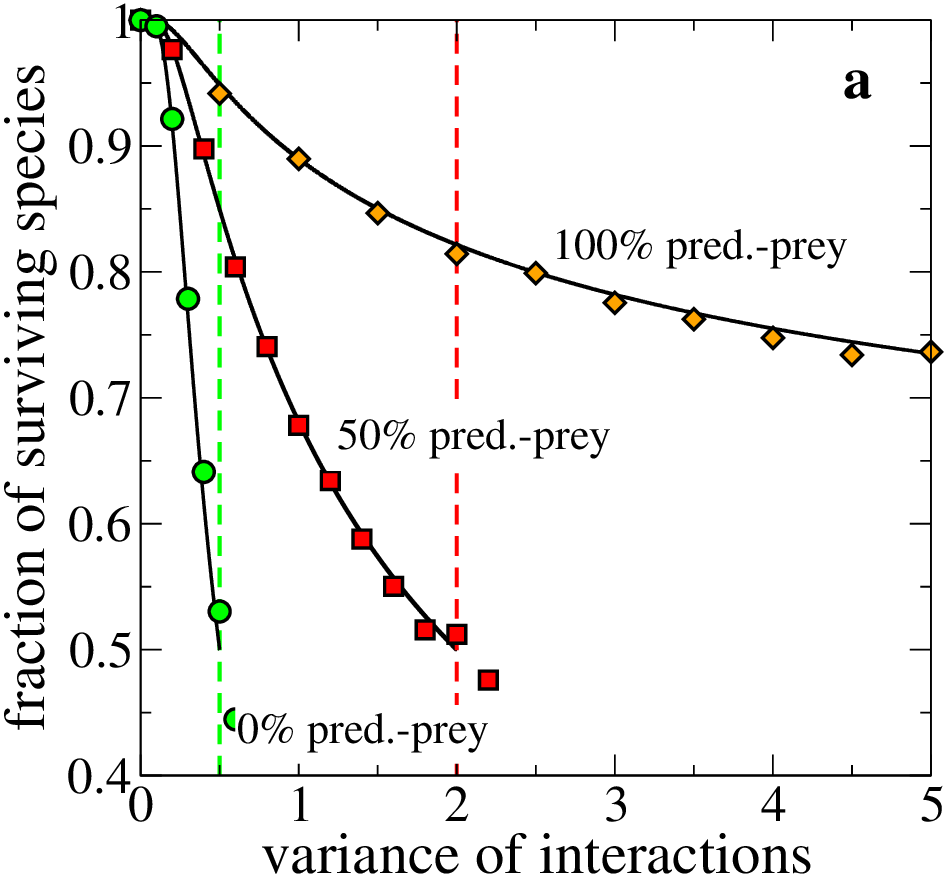}~~~\includegraphics[width=0.22\textwidth]{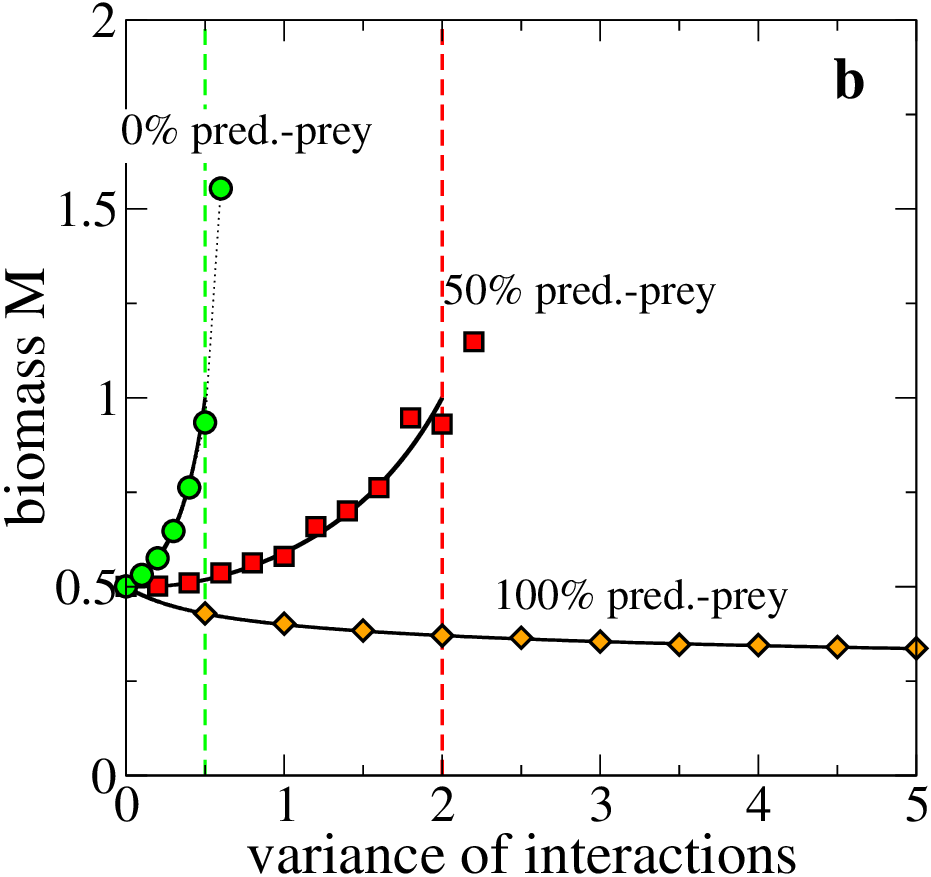}
\end{center}
% Here is how to import EPS art
\caption{\label{fig:phibiomass} (Color on-line) {\bf a} Fraction $\phi=N_S/N$ of surviving species as a function of the variance of interaction coefficients and for different percentages of predator-prey interactions in the community. Markers are data from numerical simulations ($N=300$ species, averages over $20$ runs), solid lines represent the results from the statistical physics theory. Vertical dashed lines show the border to instability for the eco-systems with $0\%$ and $50\%$ predator-prey interactions as predicted by the theory ($\sigma^2_c=0.5$ and $\sigma_c^2=2$ respectively); {\bf b} Biomass $M$ versus variance of interaction strengths.}
\end{figure}
Our approach allow us to carry the mathematical analysis of the model
further, and to investigate its properties in the stable phase. While
the Lotka-Volterra dynamics start from a community with $N$ species,
individual species may become extinct over time, and the system
may hence evolve towards a state in which fewer than $N$ species
survive asymptotically. The ratio $\phi=N_S/N$ of the number of surviving species ($N_S)$
over the number of species initially present ($N$) can be obtained from the
theory as explained above. It is depicted in Fig. \ref{fig:phibiomass}a. As seen in the
figure an excellent agreement between theoretical predictions (lines) and
results from numerical simulations (markers) is obtained, confirming the
validity of our analytical approach. The computer simulations of the Lotka-Volterra dynamics, Eq. (\ref{eq:lv}) have been carried out using a first-order Euler-forward integration scheme with dynamical time-stepping as well as a discrete-time formulation in terms of exponential functions as described for example in \cite{ives}. Both methods lead to identical results. Initial species concentrations are set to unity, $x_i(t=0)=1$ for all $i=1,\dots,N$. Results presented in all figures are for initial community sizes of typically $N=200-300$, all data is averaged over multiple ($10-200$) realizations of interaction matrices to reduce statistical errors.  

Fig. \ref{fig:phibiomass} reveals a second
central result of our analysis: the size of the eco-system in the
asymptotic state, $N_S=\phi N$, is a decreasing function of the
variance $\sigma^2$ of interaction strengths. Complexity in the
interaction matrix (as measured by $\sigma^2$) hence leads to a
reduced complexity of the remaining community of species (measured by
$N_S$). This finding is valid irrespectively of the correlation
character of the interaction matrix, i.e. independent of the
percentage of predator-prey pairs (see Fig. \ref{fig:phibiomass}a). An
increase of the complexity of interaction thus tends to destabilize
the eco-system, while at the same time reducing the size of the
food-web of survivors. Size of the remaining eco-system and stability
are thus positively correlated.  As seen above in the analytical calculation, the random community model is stable whenever more than $50$ per
cent of the initially present species survive, and unstable otherwise (see also \cite{opper}).

To illustrate the behavior of the model further we depict the biomass $M$ of the eco-system in Fig. \ref{fig:phibiomass}b, as measured by the average concentration, $M=N^{-1}\sum_i x_i$. While species diversity is reduced with increasing complexity of interactions (panel a), effects on the total biomass depend on the composition of the community, and in particular on the relative frequency of predator-prey interactions. If only few predator-prey pairs are present, biomass production is enhanced by diversity in the interaction matrix. For an ecosystem composed entirely of predator-prey pairs, however, effects of interaction strength variability are minute and confined to a small reduction of biomass generated.
\section{Feasibility}
\begin{figure}[t]
\begin{center}
\includegraphics[width=0.25\textwidth]{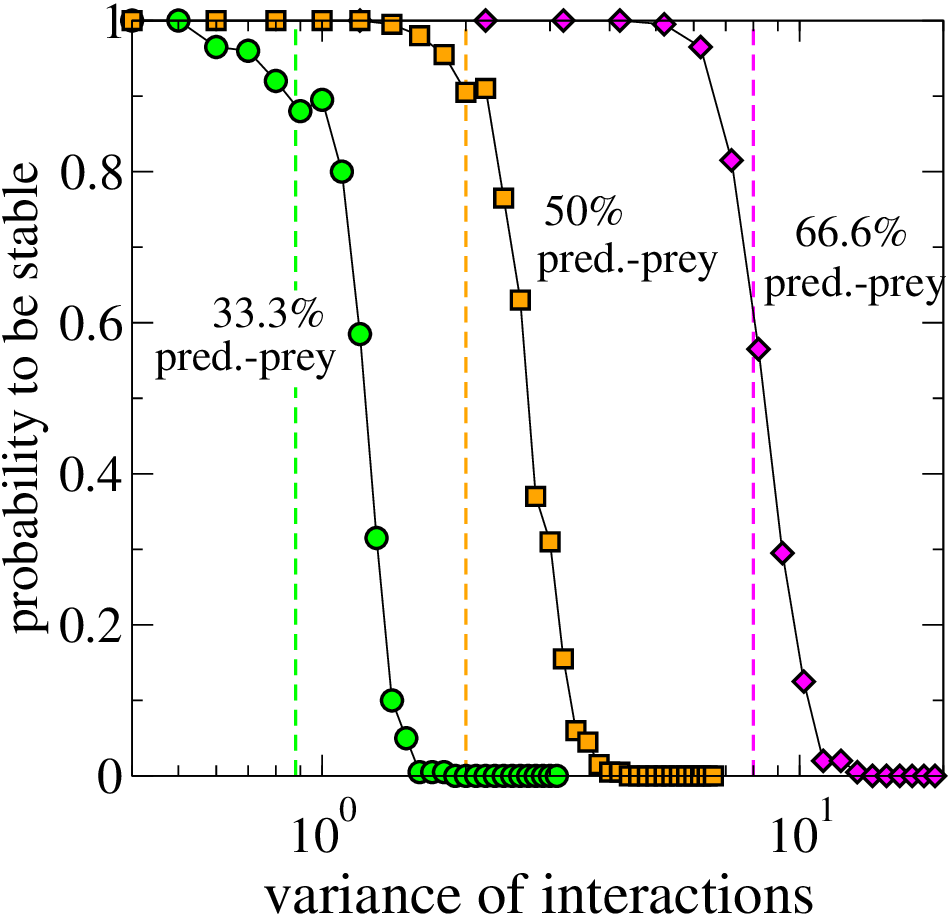}
\end{center}
% Here is how to import EPS art
\caption{\label{fig:feasibility} (Color on-line) Probability for the system to end up in a stable feasible stationary state versus the variance $\sigma^2$ of interaction matrix entries. Symbols are from simulations of communities with $N=200$ species, vertical dashed lines mark the threshold variance $\sigma_c^2$ as obtained from the theory. The system is predicted to be unstable above $\sigma_c^2$. Predictions of the theory are hence confirmed by the simulations, in which stable communities of surviving species are found for $\sigma^2<\sigma_c^2$, and where unstable behavior is observed above the threshold variance of interactions. We set $\mu=-1$, the percentage of predator-prey pairs is $33.3\%$, $50\%$ and $66.6\%$ for the left, centre and right curves respectively, corresponding to predicted values $\sigma_c^2=0.88$, $\sigma_c^2= 2$ and $\sigma_c^2=8$ of the threshold variance of interaction strengths.}
\end{figure}
Feasibility has been seen to be one of the bottlenecks limiting the
ability of species to co-exist \cite{rozdilsky}. A Lotka-Volterra
community is said to be `feasible' if all species have positive
equilibrium concentrations, and locally stable if it returns to
equilibrium after small external perturbations. We have already
examined the stability of the $N$-species Lotka-Volterra dynamics, and
now turn to its feasibility properties. To this end we have, in
numerical simulations, examined the eigenvalue properties of the
community formed by the $N_S$ survivors of the dynamics (all of which
have positive concentrations by definition). This community is subject
to a dynamics restricted to the $N_S$ non-extinct species, and gives
rise to a $N_S\times N_S$ stability matrix, of which we have obtained
the eigenvalues and stability properties numerically. In detail, labeling the $N_S=\phi N\leq N$ surviving species by $i=1,\dots,N_S$ and upon writing $x_i(t)=x_i^*+\sqrt{x_i^*}\delta_i(t)$ with $x_i^*>0$ the concentration of species $i$ at the fixed point, and with $\delta_i(t)$ a small fluctuation, a linearisation of the Lotka-Volterra dynamics leads to
\be
\frac{d}{dt}\delta_i(t)=\sum_{j=1}^{N_S} S_{ij}\delta_j(t)
\ee
with $S_{ij}=\sqrt{x_i^*}\alpha_{ij}\sqrt{x_j^*}$. See \cite{andrea} for a similar calculation. The stability of
the community of surviving species is hence governed by the
eigenvalues of the $N_S\times N_S$ stability matrix $S$. To analyze it, we have first integrated the Lotka-Volterra
dynamics, and have then identified surviving species. For each sample
generated we have then numerically computed the eigenvalues of the
so-obtained stability matrix $S$. A feasible sample is then identified
as stable if the real parts of all $N_S$ eigenvalues of $S$ are
negative.

Results are shown in
Fig. \ref{fig:feasibility}. The data confirms that the community of
survivors is robust against perturbations throughout the stable phase
predicted by the path-integral theory. A feasible stable community
hence exists for $\sigma^2<\sigma_c^2$. Above the threshold value
$\sigma_c^2$ of interaction strengths the community of survivors is
unstable, and there is no well-defined equilibrium state of the
system, but instead persistent exponential growth is found, and the
stability matrix is characterized by a positive real eigenvalue.

To generate the data of Fig. \ref{fig:feasibility}, simulations have been stopped in the unstable phase once the total asymptotically diverging biomass $M$ exceeded a threshold of the order of $10^5$. Such samples are identified as unstable.  Extinction of species in Eq. (\ref{eq:lv}) occurs exponentially, species hence become extinct only asymptotically at infinite time. Surviving species in simulations are identified as those for which $x_i(t_f)>\vartheta$, where $t_f$ denotes the time up to which the integration was performed. The threshold is chosen as $\vartheta=0.01$.
 
 Using results from random matrix theory \cite{mehta, sommers} and neglecting
correlations between $x_i^*$ and the $\{\alpha_{ij}\}$ the relevant
eigenvalue of $S$ can be identified analytically as
$\lambda_{\rm max}=-1+\sqrt{\phi}\sigma(1+\gamma)$. The stability
condition hence reads $\sqrt{\phi}\sigma<1/(1+\gamma)$. Since the
generating functional analysis reveals that $\phi=1/2$ at the onset of
instability, one recovers the above condition (\ref{eq:stab}). Note
that random matrix theory alone is not sufficient to determine
$\sigma_c^2$ as given in Eq. (\ref{eq:stab}), as knowledge of the
precise functional dependence of $\phi$ on the model parameters $\sigma,\mu,\gamma$ is
required. To our knowledge the path-integral method as sketched above is the only available analytical tool which allows one to calculate $\phi(\sigma,\mu,\gamma)$.\\

\section{Species and rank abundance}

\begin{figure}[t]
\begin{center}
 \includegraphics[width=0.25\textwidth]{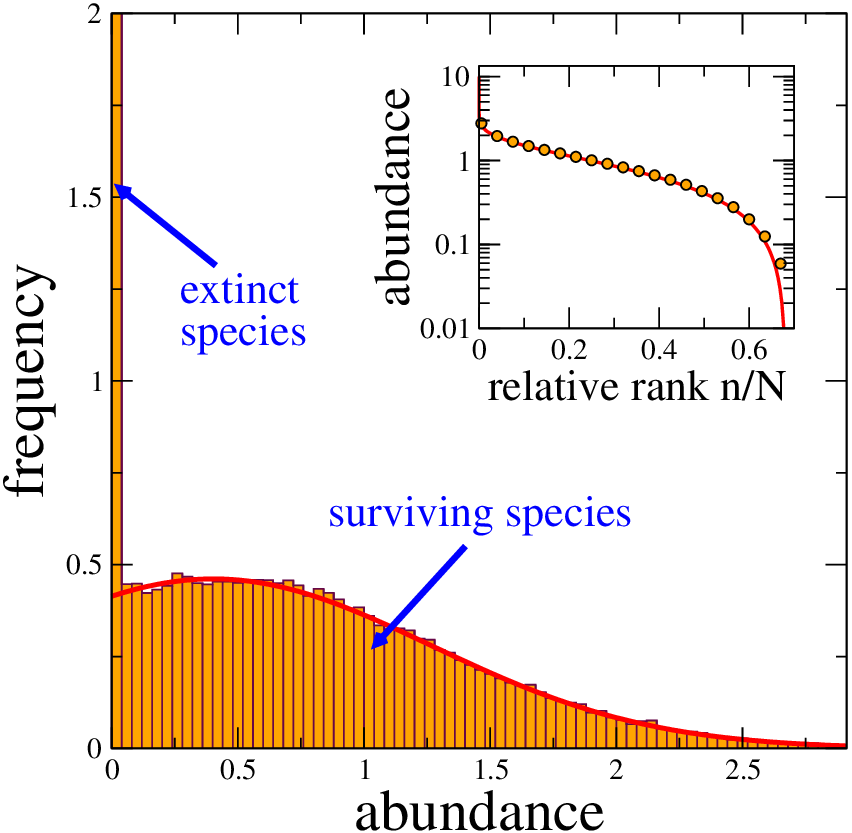}
 \end{center}
% Here is how to import EPS art
\caption{\label{fig:sad} (Color on-line) Species abundance and rank abundance distributions. The main panel shows the distribution of species concentrations. Solid line is from the statistical mechanics theory, the histogram depicted by vertical bars and shaded area from numerical simulations. Excellent quantitative agreement between theory and numerical experiment is observed. As predicted by the fixed-point theory a certain percentage of species ($31.9\%$ for the model parameters chosen in this figure) dies out asymptotically. In simulations these are found at small asymptotic densities (left-most bin). Parameters are $\mu=-1, \sigma=1$, $\alpha_{ij}$ and $\alpha_{ji}$ drawn independently for each pair $i\neq j$, i.e. $50\%$ of all species pairs are of the predator-prey type (see Supplementary Information). The inset shows the corresponding rank-abundance distribution (solid line from theory, markers from simulations). Species are ordered according to descending concentration, $x_1\geq x_2\geq...\geq x_N$, the plot shows abundance of the $n$-th species in this ranking as a function of the relative rank $n/N$. Agreement between numerical experiment (markers) and simulation (solid line) is again excellent. Numerical simulations are performed at $N=200$, averaged over $400$ samples.}
\end{figure}

The statistical mechanics theory is also able to predict
species-abundance and rank abundance distributions. This was first carried out  for the case of replicator models with symmetric random interaction matrices based on equilibrium techniques in \cite{tokita1,tokita2}, and subsequently extended to the non-equilibrium case of couplings of a general asymmetry in \cite{yoshino2}.  The off-equilibrium path-integral technique used in this paper can be used to calculate species and rank abundance for the Lotka-Volterra model. As opposed to the case of replicator models the overall biomass (closely related to the average concentration of individuals per species) is not held constant, but a dynamical property of the model.

The fraction of survivors as well as the distribution of
concentrations of the surviving species can be computed from our
analysis in the limit of large system size, without making any
approximations at any stage and compares excellently with results from
numerical simulations of systems with $N=200$ species
(Fig. \ref{fig:sad}). Our results thus improve on the analysis in \cite{wilson}, who computed abundance
relations via so called `target concentrations'. The latter may a
priori come out negative, and to circumvent this technical problem
Wilson et al. applied a heuristic cut-off, for which there is no need
in our exact approach. Still results reported in Fig. \ref{fig:sad}
are qualitatively similar to those shown in \cite{wilson}  (see e.g. their Figure 1). For the present model with uniform
carrying capacities across all species ($K_i=1$ for all
$i=1\,\dots,N$), the abundance distribution is found to be of a
Gaussian shape restricted to the positive axis. However,
generalization to species-dependent carrying capacities $K_i$ is
straightforward and inherently non-Gaussian
species-abundance relations are then to be expected.
\section{Discussion}
We have shown how concepts from theoretical spin glass physics and
disordered systems theory reveal the combined effects of asymmetric
interactions, predator-prey pairs and interaction strength variability
on the behavior of random community Lotka-Volterra models. As a key finding our
analysis provides mathematical evidence that predator-prey have a
stabilizing effect on random community Lotka-Volterra
dynamics, whereas increased variability of the inter-species interaction coefficients generally reduces stability. At the same time, increasing the complexity of couplings leads to smaller asymptotic foodwebs (due to extinction of species in the transient dynamics). Communities with a large number of surviving species are hence more likely to be stable than smaller ones.

Further application of the methods used here to related models of
theoretical ecology might provide mathematical underpinning
of some central issues of the diversity-stability debate. Studies of random
community models with heterogeneous species properties (e.g. species-specific carrying capacities), compartmental
structure \cite{rozdilsky2} or more complex interaction
graphs \cite{montoya} and dynamically evolving topologies will be
envisaged and may allow progress toward a further understanding how
complexity affects diversity and stability of ecological systems, and
what properties of the underlying interaction matrix and foodweb topology are crucial to sustain diversity. The path-integral approach goes beyond analysing random Jacobian or community matrices, and allows one to study the stability and diversity of dynamically generated fixed points in random ecosystems, including the extinction of species.  Therefore, we think that these methods are able to make useful contributions to realising for example the research programme on random community models outlined in \cite{allesina2015}.

\section{Acknowledgments}
 This work was supported by a Research Councils UK Fellowship (RCUK reference EP/E500048/1).

 \onecolumn
 \renewcommand{\theequation}{S\arabic{equation}}
 \setcounter{equation}{0}
 \centerline{{\bf ---~Supplementary Material~---}}
 \section{Generating functional analysis} 
The calculation is based on the principles of \cite{dominicis,msr,coolendyn,coolenbook1,coolenbook}. It was originally developed in the context of random replicator models in \cite{opper}, and then used for example also in \cite{gallaasym,gallahebb,gallafarmer}.
\bigskip

The dynamical generating functional is defined as
\be\label{eq:gf}
Z[\boldpsi]=\int D[\bx] \delta(\mbox{equations of motion})e^{i\sum_i\int dt~ x_i(t)\psi_i(t) }.
\ee
The source field $\boldpsi$ generates correlation functions, and will eventually be set to zero at the end of the calculation. The notation $\delta(\mbox{equations of motion})$ indicates that the integral in Eq. (\ref{eq:gf}) is over paths of the dynamics of the Lotka-Volterra equations.

Expressing the delta-functions as Fourier transforms, we have
\BE
Z[\boldpsi]
&=&\int D[\bx,\widehat\bx] \exp\Bigg(i\sum_i\int dt \Bigg[\widehat x_i(t)\Bigg(\frac{\dot x_i(t)}{x_i(t)}-[1-x(t)+\sum_{j\neq i}\alpha_{ij} x_j+h(t)]\Bigg)\Bigg]\nonumber\\
&&\times \exp\left(i\sum_i\int dt~ x_i(t)\psi_i(t)\right).\label{eq:gf0}
\EE
Next, we look at the terms containing the disorder (the $\{a_{ij}\}$), and perform the Gaussian average over these random variables, keeping in mind that for any pair $i<j$ of species we have
\be
\alpha_{ij}=\frac{\mu}{N}+\frac{\sigma}{\sqrt{N}}z_{ij}, ~~~ \alpha_{ji}=\frac{\mu}{N}+\frac{\sigma}{\sqrt{N}}z_{ji}, 
\ee
where $z_{ij}$ and $z_{ji}$ are drawn from a Gaussian distribution with $\overline{z_{ij}}=0$, $\overline{z_{ij}^2}=1$, and $\overline{z_{ij}z_{ji}}=\gamma$. 

We find, to leading order in $N$,
\BE
&&\overline{\exp\left(-i\sum_{i\neq j}\int dt \, \widehat x_i(t)a_{ij}x_j(t)\right)}\nonumber \\
&=&\exp\left(-\mu N  \int dt \, P(t) M(t)\right)\nonumber \\
&&\times\exp\left(-\frac{1}{2}N\sigma^2\int dt ~dt' \left[L(t,t')C(t,t')+\gamma K(t,t')K(t',t)\right]\right),
\EE
where we have introduced the short-hands
\BE
&M(t)=\frac{1}{N}\sum_i x_i(t),\nonumber\\
&P(t)=i\frac{1}{N}\sum_i  \widehat x_i(t),\nonumber\\
&C(t,t')=\frac{1}{N}\sum_i x_i(t) x_i(t'), \nonumber \\
&K(t,t')=\frac{1}{N}\sum_i x_i(t) \widehat x_i(t'),  \nonumber \\
&L(t,t')=\frac{1}{N}\sum_i \widehat x_i(t)\widehat x_i(t'). \label{eq:op}
\EE
These quantities can formally be introduced into the generating functional as delta-functions in their integral representation, e.g.
\BE
1&=&\int D[C]\prod_{t,t'}\delta\left(C(t,t')-\frac{1}{N}\sum_i x_i(t)x_i(t')\right)\nonumber \\
&=&\int D[\widehat C, C]\exp\left(iN\int dt~ dt' \widehat C(t,t') \left( C(t,t')-N^{-1} \sum_i x_i(t) x_i(t')\right)\right),
\EE
and similarly for the other order parameters. We have chosen the scaling of the conjugate parameter $\widehat C(t,t')$ such that the overall exponent contains a prefactor $N$. 
\\

The disorder-averaged generating functional can be written as follows
\be\label{eq:sp}
\overline{Z[\boldpsi]}=\int D[M,C,L,K,P,\widehat M,\widehat C,\widehat L,\widehat K,\widehat P]\exp\left(N\left[\Psi+\Phi+\Omega+{\cal O}(N^{-1})\right]\right).
\ee
The term
\BE
\Psi&=&i\int dt~ [\widehat M(t) M(t) +\widehat P(t) P(t)] \nonumber \\
&&+i\int dt ~dt' \left[\widehat C(t,t')C(t,t')+\widehat K(t,t')K(t,t')+\widehat L(t,t')L(t,t')\right]
\EE
results from the introduction of the macroscopic order parameters. The contribution
\BE
\Phi&=&-\frac{1}{2}\sigma^2\int dt ~dt' \left[L(t,t')C(t,t')+\gamma K(t,t')K(t',t)\right]\nonumber \\
&&-\mu\int dt ~ M(t) P(t)
\EE
comes from the disorder average, and $\Omega$ describes the details of the microscopic time evolution
\BE
\Omega&=&N^{-1}\sum_i\log\bigg[\int D[x_i,\widehat x_i] p_{0}^{(i)}(x_i(0))\exp\left(i\int dt~ \psi_i(t)x_i(t)\right)\nonumber \\
&&\times \exp\left(i\int dt ~\widehat x_i(t)  \left(\frac{\dot x_i(t)}{x_i(t)}-[1-x(t)]-h(t)\right)\right)\nonumber \\
&&\times \exp\left(-i\int dt ~ dt' \left[\widehat C(t,t')x_i(t)x_i(t')+\widehat L(t,t')\widehat x_i(t)\widehat x_i(t')+\widehat K(t,t') x_i(t)\widehat x_i(t')\right]\right)\bigg]\nonumber \\
&&\times \exp\left(-i\int dt~  [\widehat M(t)x_i(t)+\widehat P(t) i\widehat x_i(t)]\right).
\label{eq:omega}
\EE
The quantity $p_{0}^{(i)}(\cdot)$ describes the distribution from which the initial values of the $\{x_i\}$ are drawn.
\\

We next use the saddle-point method to carry out the integrals in Eq. (\ref{eq:sp}). This is valid in the limit $N\to\infty$, and amounts to finding the extrema of the term in the exponent. Setting the variation with respect to the integration variables $M, P, C,K$ and $L$ to zero gives
\BE
i\widehat M(t)&=&\mu P(t), \nonumber \\
i\widehat P(t)&=&\mu M(t), \nonumber \\
i\widehat C(t,t')&=&\frac{1}{2}\sigma^2L(t,t'),\nonumber \\
i\widehat K(t,t')&=&\gamma \sigma^2 K(t',t),\nonumber \\
i\widehat L(t,t')&=&\frac{1}{2}\sigma^2C(t,t').
\EE

Next we extremise with respect to $\widehat M, \widehat P, \widehat C, \widehat K,\widehat L$. We find
\BE
M(t)&=&\lim_{N\to\infty}N^{-1}\sum_i\avg{x_i(t)}_\Omega, \nonumber \\
P(t)&=&\lim_{N\to\infty}N^{-1}\sum_i\avg{i\widehat x_i(t)}_\Omega, \nonumber \\
C(t,t')&=&\lim_{N\to\infty}N^{-1}\sum_i\avg{x_i(t)x_i(t')}_\Omega,\nonumber \\
K(t,t')&=&\lim_{N\to\infty}N^{-1}\sum_i\avg{x_i(t)\widehat x_i(t')}_\Omega, \nonumber \\
L(t,t')&=&\lim_{N\to\infty}N^{-1}\sum_i\avg{\widehat x_i(t)\widehat x_i(t')}_\Omega, 
\EE
where the average $\avg{\dots}_\Omega$ is to be taken against a measure defined by the exponent of the expression in Eq. (\ref{eq:omega}) in the limit $h\to 0$, see e.g. \cite{opper,coolendyn, coolenbook1,coolenbook, gallaasym} for similar calculations. 

From Eq. (\ref{eq:gf0}) (and taking the thermodynamic limit) one also notices that
\BE
C(t,t')&=&-\lim_{N\to\infty}N^{-1}\sum_i\left.\frac{\delta ^2\overline{Z[\boldpsi]}}{\delta \psi_i(t)\delta\psi_i(t')}\right|_{\boldpsi=0, h=0}, \nonumber \\
K(t,t')&=&\lim_{N\to\infty}N^{-1}\sum_i\left.\frac{\delta ^2 \overline{Z[\boldpsi]}}{\delta \psi_i(t)\delta h(t')}\right|_{\boldpsi=0, h=0}, \nonumber \\
 L(t,t')&=&-\lim_{N\to\infty}N^{-1}\sum_i\left.\frac{\delta ^2  \overline{Z[\boldpsi]}}{\delta h(t)\delta h(t')}\right|_{\boldpsi=0, h=0},\nonumber\\
 P(t)&=&-\lim_{N\to\infty}N^{-1}\sum_i\left.\frac{\delta\overline{Z[\boldpsi]}}{\delta h(t)}\right|_{\boldpsi=0, h=0}, 
\EE

Given that $Z[\boldpsi=0,h]=1$ for all $h$ due to normalisation we conclude that $L(t,t')=0$ for all $t,t'$, and $P(t)=0$ for all $t$. We now set $\boldpsi=0$. We will also assume that initial conditions are chosen from identical distributions for all components $x_i$  (i.e. $p_{0}^{(i)}(\cdot)$ does not depend on $i$). Then we have
\BE
\Omega&=&\log\bigg[\int D[x,\widehat x]~ p_{0}(x(0)) \exp\left(i\int dt ~\widehat x(t)  \left(\frac{\dot x(t)}{x(t)}-[1-x(t)]-h(t)-\mu M(t)\right)\right)\nonumber \\
&&\times \exp\left(-\sigma^2\int dt ~ dt' \left[ \frac{1}{2}C(t,t')\widehat x(t)\widehat x(t')+i\gamma G(t',t) x(t)\widehat x(t')\right]\right)\bigg]\label{eq:omega2}
\EE
where we have used the above saddle-point results, and where we have introduced $G(t,t')=-iK(t,t')$.

The final result for the generating functional post disorder average is therefore
\BE
Z_{\mbox{\footnotesize eff}}&=&\int D[x,\widehat x] ~p_{0}(x(0)) \exp\left(i\int dt ~\widehat x(t)  \left(\frac{\dot x(t)}{x(t)}-[1-x(t)]-\mu M(t)-h(t)\right)\right)\nonumber \\
&&\times \exp\left(-\sigma^2\int dt ~ dt' \left[ \frac{1}{2}C(t,t')\widehat x(t)\widehat x(t')+i\gamma G(t',t) x(t)\widehat x(t')\right]\right).
\EE
This is is recognised as the generating function of the {\em effective} dynamics
\BE
\dot x(t)=x(t)\left[1-x(t)+\gamma \sigma^2\int dt' G(t,t') x(t')+\mu M(t)+\eta(t)+h(t)\right], \label{eq:effective_s}
\EE
where
\BE
 G(t,t')&=&\avg{\frac{\delta x(t)}{\delta h(t')}}_*,\nonumber \\
 \avg{\eta(t)\eta(t')}_*&=&\sigma^2\avg{x(t)x(t')}_*,\nonumber \\
 \avg{x(t)}_*&=&M(t),
\EE
and where $\avg{\cdots}_*$ denotes an average over realizations of the effective dynamics (\ref{eq:effective_s}). Given that this is to be evaluated at $h=0$ we can equivalently write
\BE
\dot x(t)=x(t)\left[1-x(t)+\gamma \sigma^2\int dt' G(t,t') x(t')+\mu M(t)+\eta(t)\right], \label{eq:effective2}
\EE
with
\BE
 G(t,t')&=&\avg{\frac{\delta x(t)}{\delta \eta(t')}}_*,\nonumber \\
 \avg{\eta(t)\eta(t')}_*&=&\sigma^2\avg{x(t)x(t')}_*,\nonumber \\
 \avg{x(t)}_*&=&M(t). \label{eq:selfc}
\EE
Eqs. (\ref{eq:effective2}) and (\ref{eq:selfc}) determine $G(t,t'), C(t,t')=\avg{x(t)x(t')}$ and $M(t)$ self-consistently.

\section{Fixed point analysis}\label{subsub4}

We now assume that the system reaches a stationary state and that this stationary state does not depend on the initial condition (i.e., we assume absence of long-term memory, see also \cite{gallaasym}). The response function $G$ is then a function of time differences only, i.e. $G(t,t')=G(\tau)$, where $\tau=t-t'$. Causality dictates $G(\tau<0)=0$. Assuming further that the dynamics reaches a fixed point, $C(t,t')$ is constant (independent of $t$ and $t'$); we write $C(t,t')\equiv q$. 

Fixed points of the effective dynamics are given by the solutions of 
\be \label{eq:fp}
x^*\left[1-x^*+\gamma\sigma^2\chi x^*+\mu M^*+\eta^*\right]=0,
\ee
where we have written $\chi=\int_0^\infty d\tau~ G_(\tau)$. We note that $\eta(t)$ becomes {\em static} Gaussian randomness, $\eta^*$, at the fixed point, due to the self-consistency relation $ \avg{\eta(t)\eta(t')}_*=\sigma^2\avg{x(t)x(t')}_*\equiv \sigma^2 q$. We write $\eta^*=\sqrt{q}\sigma z$ with $z$ a static Gaussian random variable of mean zero and unit variance. 
\\

Eq. (\ref{eq:fp}) always has the solution $x^*=0$. The second solution, 
\be
x^*=\frac{1+\mu M^*+\sqrt{q}\sigma z}{1-\gamma\sigma^2\chi}
\ee
is physical when this expression is non-negative. In the following we use
\be\label{eq:theta}
x(z)= \frac{1+\mu M^*+\sqrt{q}\sigma z}{1-\gamma\sigma^2\chi}H\left(\frac{1+\mu M^*+\sqrt{q}\sigma z}{1-\gamma\sigma^2\chi}\right),
\ee
where $H(x)$ is the Heaviside function, $H(x)=1$ for $x>0$, and $H(x)=0$ else. The zero solution can be seen to be unstable when the expression in the Heaviside function is positive, see below.

The order parameters $\chi$, $q$ and $M$ are to be determined from the self-consistency relations
\BE
\chi&=&\frac{1}{\sqrt{q}\sigma}\avg{\frac{\partial x(z)}{\partial z}}_*,\nonumber \\
\avg{x(z)}_*&=&M^*,\nonumber\\
q&=&\avg{(x(z))^2}_*.
\EE
This can be expressed as follows
\BE
\chi&=&\frac{1}{\sqrt{q}\sigma}\int_{-\infty}^\infty Dz ~\frac{\partial x(z)}{\partial z}, \nonumber \\
M^*&=&\int_{-\infty}^\infty Dz~x(z),\nonumber \\
q&=&\int_{-\infty}^\infty Dz~ x(z)^2, \label{eq:fpsc}
\EE
where $Dz=\frac{dz}{\sqrt{2\pi}}e^{-z^2/2}$. 

Only the non-zero fixed points contribute to these integrals. We proceed under the assumption $1-\gamma\sigma^2\chi>0$ (see below for further discussion). The range $x(z)>0$ is then equivalent to $1+\mu M^*+\sqrt{q}\sigma z>0$, i.e. $z>-\Delta$, where $\Delta=(1+\mu M^*)/(\sqrt{q}\sigma)$. This means that the fraction of surviving species is given by $\phi=\int_{-\Delta}^\infty Dz$, which --- due to symmetry of the Gaussian integrand --- can also be written as $\phi = \int_{-\infty}^\Delta Dz$. In the integration range we have
\be
x(z)=\sqrt{q}\sigma \frac{\Delta+z}{1-\gamma\sigma^2\chi}
\ee
Eqs. (\ref{eq:fpsc}) then turn into
\BE
\chi&=&\frac{1}{1-\gamma\sigma^2\chi}\int_{-\Delta}^\infty Dz, \nonumber \\
M^*&=&\sqrt{q}\sigma \frac{1}{1-\gamma\sigma^2\chi}\int_{-\Delta}^\infty Dz~ (\Delta+z),\nonumber\\
1&=&\frac{\sigma^2}{(1-\gamma\sigma^2\chi)^2}\int_{-\Delta}^\infty Dz~ (\Delta+z)^2.
\EE
Changing the integration variable $z$ into $-z$ this is
\BE\label{eq:sc3}
\chi&=&\frac{1}{1-\gamma\sigma^2\chi}\int_{-\infty}^\Delta Dz, \nonumber \\
M^*&=&\sqrt{q}\sigma \frac{1}{1-\gamma\sigma^2\chi}\int_{-\infty}^\Delta Dz ~(\Delta-z),\nonumber\\
1&=&\frac{\sigma^2}{(1-\gamma\sigma^2\chi)^2}\int_{-\infty}^\Delta Dz~ (\Delta-z)^2.
\EE
 These are the expressions given in Eq. (9-11) of the main paper.
 
 Along the way, we have made the assumption $1-\gamma\sigma^2\chi>0$. This can be checked retrospectively from the numerical solution. It is also required self-consistently in the third relation in Eq. (\ref{eq:sc3}), as $M^*\geq0$. We also note that the first relation in Eq. (\ref{eq:sc3}) then implies $\chi>0$, which we will use below.
 \section{Linear stability analysis}

We proceed along the lines of \cite{opper}. We add white noise $\xi(t)$ of unit variance to the effective process
\BE
\dot x(t)=x(t)\left[1-x(t)+\gamma \sigma^2\int dt' G(t,t') x(t')+\mu M(t)+\eta(t)+\varepsilon\xi(t)\right]. \label{eq:effective22}
\EE
We study fluctuations $y(t)$ about a fixed point of Eq. (\ref{eq:effective2}), i.e. we write $x(t)=x^*+\varepsilon y(t)$, and denote the resulting additional term in the self-consistent noise by $\varepsilon v(t)$ (i.e., $\eta(t)=\eta^*+\varepsilon v(t)$). We use $\avg{\xi(t)}=0$, $\avg{v(t)}=0$ and $\avg{y(t)}=0$. We linearise in $\varepsilon$, i.e. in $y(t), v(t)$ and $\xi(t)$.
\medskip

We first consider the case $x^*=0$. Linearising the effective process one has
\be
\frac{dy(t)}{dt}=y(t)\left[1+\mu M^*+\sqrt{q}\sigma z\right].
\ee
Within our ansatz, the object in the square brackets is negative for fixed points at zero [see Eq. (\ref{eq:theta}), and noting again that $1-\gamma\sigma^2\chi>0$]. We conclude that perturbations around zero fixed points decay. We also note that converseley the zero fixed point is not stable if the object in the square brackets is positive, justifying retrospectively that we use the non-zero solution $x^*$ in this case --- see again Eq. (\ref{eq:theta})].

For non-zero $x^*$ we have, to linear order in $y, v$ and $\xi$,
\BE
\frac{dy(t)}{dt}&=&x^*\left[-y(t)+\gamma\sigma^2\int_{0}^t G(t,t')y(t')dt'+v(t)+\xi(t)\right]. \label{eq:pertt}
\EE
Self-consistently one has $\avg{v(t)v(t')}=\sigma^2\avg{y(t)y(t')}$. One converts into Fourier space and obtains
\BE
\frac{i\omega\widetilde y(\omega)}{x^*}&=&\left(\gamma\sigma^2 \widetilde G(\omega)-1\right) \widetilde y(\omega)+\widetilde v(\omega)+\widetilde \xi(\omega). \label{eq:pertt2}
\EE
This leads to
\be\label{eq:help7}
\left[\frac{i\omega}{x^*}+\left(1-\gamma\sigma^2 \widetilde G(\omega)\right)\right] \widetilde y(\omega)=\widetilde v(\omega)+\widetilde \xi(\omega), 
\ee
We note that $\avg{|\widetilde y(\omega)|^2}$ is the Fourier transform of $\avg{y(t)y(t+\tau)}$, assuming a stationary state in which this correlation function only depends on $\tau$. Then $\avg{|\widetilde y(\omega=0)|^2}=\int d\tau\, \avg{y(t)y(t+\tau)}$. If this quantity diverges, perturbations do not decay to zero, signalling an instability of the fixed point. Hence we focus on $\omega=0$. Using $\widetilde G(\omega=0)=\chi$ and $\avg{|\widetilde v(0)|^2}=\sigma^2\avg{|\widetilde y(0)|^2}$ this leads to
\be
\avg{|\widetilde y(0)|^2}=\phi \left[\gamma\sigma^2 \chi-1\right]^{-2}\left[\sigma^2\avg{|\widetilde y(0)|^2} +1\right]. \label{eq:pertt3}
\ee
The factor $\phi$ on the right-hand side arises because Eq. (\ref{eq:help7}) only applies to non-zero fixed points (fluctuations about zero fixed points decay, as demonstrated above, and so they do not contribute to $\avg{|\widetilde y(\omega=0)|^2}$).

Eq. (\ref{eq:pertt3}) can be re-written as
\be
\avg{|\widetilde y(0)|^2}=\frac{\phi}{\left[\gamma\sigma^2 \chi-1\right]^2-\phi\sigma^2}.
\ee
This indicates that $\avg{|\widetilde y(0)|^2}$ diverges when $\phi\sigma^2=(1-\gamma\sigma^2\chi)^2$.  One finds $\phi\sigma^2<(1-\gamma\sigma^2\chi)^2$ in the stable phase, consistent with a well-defined (positive) quantity $\avg{|\widetilde y(0)|^2}$.  

The condition $\phi\sigma^2=(1-\gamma\sigma^2\chi)^2$ leads to $\Delta=0$ in Eqs. (9-11) of the main paper. To see this we  insert this condition into $1=\frac{\sigma^2}{(1-\gamma\sigma^2\chi)^2}\int_{-\infty}^\Delta Dz (\Delta-z)^2$ and find
\be
\phi=\int_{-\infty}^\Delta Dz~ (\Delta -z)^2.
\ee
On the other hand we also have $\phi=\int_{-\infty}^\Delta Dz$. Comparing the two expressions gives $\Delta=0$, and hence $\phi=1/2$.

Using this, we have
\BE\label{eq:helppp}
2\chi&=&\frac{1}{1-\gamma\sigma^2\chi}, \nonumber \\
2&=& \frac{\sigma^2}{(1-\gamma\sigma^2\chi)^2},
\EE
from which we find $\chi^2=1/(2\sigma^2)$. Using $\chi>0$ we have $\chi=1/(\sqrt{2}\sigma)$. Substituting this in the first relation in Eq. (\ref{eq:helppp}) in turn leads to
 
\be\label{eq:stabb}
\sigma_c^2(\gamma)=\frac{2}{(1+\gamma)^2}.
\ee

\end{document}